\documentclass[10pt,conference]{IEEEtran}
\usepackage{array}
\usepackage{cite}
\usepackage{calc}
\usepackage{caption}
\usepackage{graphicx}
\usepackage{psfrag}
\usepackage[tight,footnotesize]{subfigure}
\usepackage{stfloats}
\usepackage{url}
\usepackage{subfig}
\usepackage{longtable}
\usepackage{stfloats}
\usepackage{amsfonts,amssymb,amsbsy,bm,paralist,theorem,cite,ifthen,color}
\usepackage[cmex10]{amsmath}
\usepackage{algorithmic,algorithm}

\newcommand\xb{\ensuremath{{\bm x}}}

\newcommand\ssb{\ensuremath{{\bm s}}}

\newcommand\db{\ensuremath{{\bm d}}}

\newcommand\Eb{\ensuremath{{\bf E}}}

\newcommand\zb{\ensuremath{{\bm z}}}
\newcommand\mub{\ensuremath{{\bm \mu}}}
\newcommand\lambdab{\ensuremath{{\bm \lambda}}}

\newcommand\Omegab{\ensuremath{{\bm \Omega}}}
\newcommand\Upsilonb{\ensuremath{{\bm \Upsilon}}}
\newcommand\zerob{\ensuremath{{\bm 0}}}

\newcommand\Psib{\ensuremath{{\bf \Psi}}}

\newcommand\E{\ensuremath{{\mathbb{E}}}}

\newcommand\pib{\ensuremath{{\bm \pi}}}

\begin{document}
\bibliographystyle{IEEEtran}
\title{Coordinated Home Energy Management for Real-Time Power Balancing}

\author{Tsung-Hui Chang, Mahnoosh Alizadeh, and Anna Scaglione \\ ~ \\
\begin{tabular}{cc}
Department of Electrical and Computer Engineering    \\
University of California, Davis,   \\
One Shields Avenue, Davis, California 95616 \\
\small E-mail: tsunghui.chang@ieee.org, malizadeh@ucdavis.edu, ascaglione@ucdavis.edu
\end{tabular}
\vspace{-\baselineskip}}

\maketitle
\begin{abstract}
This paper proposes a coordinated home energy management system (HEMS) architecture where the distributed residential units cooperate with each other to achieve real-time power balancing. The economic benefits for the retailer and incentives for the customers to participate in the proposed coordinated HEMS program are given. We formulate the coordinated HEMS design problem as a dynamic programming (DP) and use approximate DP approaches to efficiently handle the design problem. A distributed implementation algorithm based on the convex optimization based dual decomposition technique is also presented. Our focus in the current paper is on the deferrable appliances, such as Plug-in (Hybrid) Electric Vehicles (PHEV), in view of their higher impact on the grid stability. Simulation results shows that the proposed coordinated HEMS architecture can efficiently improve the real-time power balancing.
\end{abstract}

\IEEEpeerreviewmaketitle
\vspace{-0.1cm}
\section{Introduction}
Even though Demand Response (DR) \cite{dr} technologies were studied and practiced since the 60s, their integration in the US wholesale markets has been facilitated by several regulatory rules and measures ever since the state of California was struck by energy crises in 2000 and 2001. Available DR technologies are mainly categorized into the following: Direct Load Control (DLC) strategies \cite{dlc1,dlc2}, where a controller centrally interrupts the jobs of participating appliances mostly in case of emergencies and to curtail high peak load; Dynamic Pricing programs \cite{Born}, which includes several rates and tariffs to manage the demand for electricity in a decentralized manner, e.g., Time of Use (TOU), Critical Peak Pricing (CPP), Real Time Pricing (RTP) and Day Ahead Pricing (DAP) rates; Demand bidding programs \cite{dsb}, where a market participant 
directly makes an offer to the wholesale market (or the retailer) for reducing electricity during peak times on the next day.

All the above mentioned strategies have their own pros and cons. For example, DLC, probably the oldest and safest measure of demand management, unfortunately cannot happen frequently, and thus they can offer little flexibility for integrating intermittent renewable resources into the grid. They are mainly designed for emergencies and cannot easily account for the inconvenience they cause to their customers, i.e., the Quality of Service (QoS) provided. TOU rates are designed months in advance and cannot handle real-time load management in case of emergencies or help integrate intermittent resources into the power grid. RTP may be the most practical and probably the cheapest way of managing electricity demand in the future but it faces the challenging problem of what these price signals should be to avoid causing physical and market instabilities while reflecting the true conditions of the market at the same time. In fact, it has been shown that RTP are likely to cause more volatility or even instabilities when customers respond to this new information and form a new feedback loop in the power system control model \cite{certs,box,mitter}. 

Given dynamic pricing tariffs, the responsibility of managing demand in response to these price signals cannot be left to the consumer and should be mostly automated. Consequently, there is an extensive literature emerging on Home Energy Management
Systems (HEMS), e.g.,  \cite{han,rad}. In these works, researchers look into finding optimal designs for the software and hardware suited for residential use that would respond to these price signals in an automated fashion. These HEMS units receive requests from their owners specifying the appliances they plan to use in the near future and their preferences. The software then runs an optimization that plans the use of these appliances, based on
their power consumption, job deadlines and other customer
specified factors, taking into account the dynamic price made available to the unit from its associated retailer/aggregator.

As observed in \cite{Kishore2010}, since all the residences are given the same dynamic price,
current HEMS that individually operated by each residence will simultaneously schedule the load to the low-price period, and, consequently, a new ``rebound" peak is created to the grid.  In this paper, we aim to blur the boundaries between RTP and DLC strategies by proposing an architecture through which HEMS units inside the territory of an aggregator/ratailer can cooperate with each other to keep the demand presented by the retailer to the wholesale market balanced with the available generation supply (which might be the day-ahead bid plus locally available renewable resources).
Several exiting works have considered such a coordinated energy management architecture, though different goals are considered. For example, in \cite{Kishore2010} a heuristic neighborhood-level energy management algorithm is proposed for scheduling the load of residential units such that the aggregate load meets a maximum power profile specified by the retailer. In \cite{Mohsenian-Rad2010}, a distributed energy management algorithm, based on a game-theoretic approach, was proposed to minimize the cost of the retailer as well as the peak-to-average ratio of the aggregate load. 
The work in \cite{Gatsis2011} takes into account user dissatisfaction and proposed a distributed energy management algorithm that minimizes the cost of retailer and a cost that reflects the degree of user dissatisfaction. Both the works in \cite{Mohsenian-Rad2010} and \cite{Gatsis2011} assumed that the operating times of appliances are known \emph{a priori}, and, moreover, they allowed HEMS to optimize the load injection of appliances. However, in many cases, the appliances have fixed load profiles that cannot be altered. The operating times of appliances should also depend upon the householder's request which are usually random. Another issue that is not clearly addressed in the existing works is the incentives for the customers to participate in the energy management coordination.

Our intention in this paper is to propose a coordinated HEMS architecture where the HEMS units in the residences collaborate to minimize the cost of the aggregator/retailer in the real-time market.
Specifically, in addition to the cost for the day-ahead market, real-time power imbalances will further cost the retailer in the real-time market. 
Therefore, minimizing the real-time market cost directly achieves the goal of real-time power balancing in the grid.

The scenario under consideration is that the retailer will inform the customers the dynamic price, and the HEMS in each individual residence will optimize their electricity cost by scheduling its appliance activities according to the price. To encourage the customers to join the proposed coordinated HEMS program, we assume that the customers won't pay additional money compared to the cost they optimized using the individual HEMS. Moreover, the degree of comfort of customers (e.g., appliance scheduling deadline) will be taken into account in the coordinated HEMS architecture. Under such conditions, the retailer will directly benefit from the coordinated HEMS architecture while the customers will sustain no loss neither financially nor in their degree of comfort.

Different from \cite{Mohsenian-Rad2010} and \cite{Gatsis2011}, the current work assumes that the times for which the customer may submit a request for an appliance are random. Moreover, given the load profiles of appliances, we optimally defer their operating times so as to minimize the real-time market cost of the retailer. Such deferrable appliances include Plug-in (Hybrid) Electric Vehicles (PHEV), dish washer etc. which usually have higher impact on the grid power balancing. We show that the HEMS and proposed coordinated HEMS design problem can be formulated as a dynamic programming (DP). The approximate DP approach known as certainty equivalent control (CEC) \cite{BK:Bersekas07} is used to efficiently handle the considered design problems. Furthermore, the convex optimization based dual decomposition technique \cite{Boyddecomposition} is applied for developing a distributed implementation algorithm for the proposed coordinated HEMS. Simulation results are presented to demonstrate the effectiveness of the coordinated HEMS architecture.

\section{System Model and HEMS}
We consider a general wholesale market scenario where the retailer bids to purchase electricity from the market and serves a number of residential units. 
Each residence runs an energy management program for minimizing its electricity cost. This section presents the residential appliance load model and mathematical formulation of HEMS.



\subsection{Appliance Load Model}

Consider $N$ appliances in each residence that are controllable by the HEMS; for example, the PHEV, dish washer, washing machine, and cloth dryer etc., that are flexible in their operating time and allow the HEMS to defer their schedule within the deadline specified by the customers.
The load profiles of the controllable appliances are known and, once the appliances are ON, their operation cannot be interrupted. Our interest on deferrable appliances is mainly because the power consumption of deferrable appliances, especially PHEV, has a higher impact on gird stability. To model the deferrable load, we adopt the signal model presented in \cite{ddls2}, which, as will be seen later, can greatly simplify the appliance scheduling optimization problems encountered in HEMS and the proposed coordinated HEMS.

Let $g_i(\ell)$, $\ell=1,\ldots,G_i$, denote the discrete-time load profile\footnote{In this paper, for simplicity, we will assume only active power and ignore the reactive power of each appliance. If necessary, the reactive power can be easily incorporated into the developed algorithms in the subsequent sections.}\footnote{$g_i(\ell)=0$ for $\ell<1$ and $\ell>G_i$.} of appliance $i$ where $G_i>0$ is the maximum duration of $g_i(\ell)$, for $i=1,\ldots,N.$ Assuming that the customer sends requests for appliance $i$ at time $t_{i,1}, t_{i,2}, \ldots$ $\in \{1,\ldots,L\}$, where $L>0$ denotes the maximum time horizon.
Then, if without scheduling, the load injection due to appliance $i$ is given by
\begin{align}
   {D}_i(\ell)&=
   \sum_{k=1}^{\infty} g_i(\ell- t_{i,k}),~\ell=1,\ldots,L. 
\end{align}
One can describe the requests for appliance $i$
as a request arrival process:
\begin{align}
   a_i(\ell)&=\sum_{k=1}^{\infty} u(\ell- t_{i,k}),~ \ell=1,\ldots,L,
\end{align}
where $u(t)$ is the unit step function\footnote{$u(t)$ is equal to one for $t\geq 0$ and zero otherwise}. To model the customer's behavior in using appliance $i$, we assume that the arrival process $a_i(\ell)$ is a non-stationary random process with the average number of new arrivals at time $\ell$ being $\alpha_\ell\in [0,1]$, i.e., $\E\{a_i(\ell)-a_i(\ell-1)\}=\alpha_\ell$. For example, one may model $a_i(\ell)-a_i(\ell-1)$ as a binary random variable with $\alpha_\ell$ being the probability that appliance $i$ will be requested at time $\ell$.

The requested tasks of controllable appliances may be queued and scheduled to be ON later depending on the control of HEMS. Suppose that $s_{i,1}$, $ s_{i,2}$, $\ldots$ $\in \{1,\ldots,L\}$, are the scheduled operating times of appliance $i$, where $s_{i,k} \geq t_{i,k}$ for all $k$. Then the scheduled load injection of appliance $i$ is given by
\begin{align}\label{eq:load injection of appliance i}
   {S}_i(\ell)&=\sum_{k=1}^{\infty} g_i(\ell- s_{i,k}),~\ell=1,\ldots,L. 
\end{align} Similarly, the operating times of appliance $i$ can also be described by a task departure (launching) process as
\begin{align}\label{eq:departure process}
   d_i(\ell)&=\sum_{k=1}^{\infty} u(\ell- s_{i,k}),~ \ell=1,\ldots,L.
\end{align}

The total load injection of a residence is the summation of the controllable load and uncontrollable load (e.g., lights, stove etc.), and can be expressed as
\begin{align}\label{eq:total load}
     D_{{{\rm total}}}(\ell) = {U}(\ell) + \sum_{i=1}^NS_{i}(\ell),
\end{align}where ${U}(\ell)$ is the load of the uncontrollable appliances. 

\subsection{HEMS}

Given the dynamic electricity prices from the retailer, denoted by $p(\ell)$, $\ell=1,\ldots,L$, HEMS targets to schedule the controllable appliances such that the average total electricity cost of the residence, i.e.,
\begin{align}\label{eq:bill}
   \sum_{\ell=1}^L \E\{p(\ell)  D_{{{\rm total}}}(\ell)\}
\end{align}
can be minimized. The scheduling task is usually subject to a constraint that reflects the customer's degree of comfort. Here we assume that the customer will preassign a maximum tolerable delay for each appliance, and the HEMS has to turn on the appliance before the specified deadline. In particular, we denote $\zeta_i\geq 0$ as the maximum delay time of appliance $i$. Then the operating times of appliance $i$ have to satisfy
\begin{align}\label{eq:deadline constraint0}
  t_{i,k} \leq s_{i,k} \leq  t_{i,k}+\zeta_i,
\end{align} for all $k$, in order to fulfill the degree of comfort of the customer.

Mathematically, the HEMS design problem can be formulated as the following multi-stage stochastic optimization problem
\begin{subequations}\label{eq:HEMS}
\begin{align}
  \min_{s_{i,1},s_{i,2},\ldots}~&\sum_{\ell=1}^L \E\left\{p(\ell) \left(\sum_{i=1}^N S_{i}(\ell)\right)\right\} \\
  \text{subject to (s.t.)}~& {S}_i(\ell)=\sum_{k=1}^{\infty} g_i(\ell- s_{i,k})~\forall~i,\ell, \\
  & t_{i,k} \leq s_{i,k} \leq  t_{i,k}+\zeta_i~\forall~i,k, \\
  & s_{i,k} \leq L~ \forall~i,k, \label{eq:HEMS d}\\
  & \sum_{i=1}^N S_{i}(\ell) \leq P_{\max}~\forall~ \ell. \label{eq:HEMS e}
\end{align}
\end{subequations}
where \eqref{eq:HEMS d} implies that all the appliances have to be scheduled before the horizon $L$, and $P_{\max}$ in \eqref{eq:HEMS e} denotes the maximum power flow constraints of the residence.
As will be detailed later, problem \eqref{eq:HEMS} can be formulated as a dynamic programming (DP) problem and can be efficiently handled by approximate DP techniques \cite{BK:Bersekas07}

While the dynamic prices $\{p(\ell)\}_{\ell=1}^L$ are designed by the retailer such that the customers would move their load to the off-peak period of the power grid, as pointed out in \cite{Kishore2010}, the HEMS individually operated by each residence may create a new ``rebound" peak in the low-price period that can be even more severe than that without HEMS. 
As a result, the aggregate load injection from multiple HEMS-based residential units will not necessarily follow the energy supply scheduled by the day-ahead market, and the resultant real-time power balancing would increase the cost of the retailer in the wholesale real-time bidding market. In the next section, we propose to coordinate the energy management of multiple residences, aiming at minimizing the wholesale real-time market cost of the retailer. The benefits of such a coordinated energy management architecture, which we refer to as \emph{coordinated HEMS}, will be demonstrated via computer simulations.

\section{Coordinated HEMS}
We first analyze the costs of the retailer and the incentives to the customers so that the customers would like to join the proposed coordinated HEMS program. The mathematical formulation of the proposed coordinated HEMS will be presented in the second subsection. The third subsection shows how the coordinated HEMS design problem can be recast as a standard DP and can be handled by the approximate DP technique known as certainty equivalent control (CEC) \cite{BK:Bersekas07}.

\subsection{Cost of Retailer and Incentives to Customers}

The cost of the retailer mainly consists of two parts, namely, the wholesale day-ahead market bidding cost and the wholesale real-time market bidding cost. In the day-ahead market, the retailer bids to purchase energy from the generator through ISO according to the predicted load requirement for the upcoming day. Due to the prediction errors, the load actually consumed in real time may deviate from the scheduled energy supply. Under such circumstances, the retailer has to purchase additional amount of energy in the real-time market or pay to the grid for absorbing the extra energy that cannot be consumed, in order to maintain the real-time power balancing. Let $\pi_{\rm p}(\ell)$ be the price for buying energy from the real-time market and $\pi_{\rm s}(\ell)$ be the price for absorbing extra energy (if $\pi_{\rm s}(\ell)\leq 0$ then it implies that the retailer may sell back the extra energy). Let $E(\ell)$ be the energy supply. Moreover, assume that there are totally $M$ residential units, each of which contributes $D_{\rm total}^{(m)}$ (see \eqref{eq:total load}) load injection to the system.
The total real-time market cost of the retailer is given by
\begin{align}\label{eq:deviation cost}
   \text{Cost}_{RT}=&\sum_{\ell=1}^L \left[ \pi_{\rm s}(\ell)\left(E(\ell)-\sum_{m=1}^M D_{\rm total}^{(m)}(\ell)\right)^+ \notag \right.\\
    &\left.~~~~~~~~~~~+\pi_{\rm p}(\ell)\left(\sum_{m=1}^M D_{\rm total}^{(m)}(\ell)-E(\ell)\right)^+ \right],
\end{align} where $(x)^+=\max\{x,0\}$.
The profit of the retailer can be roughly calculated as
\begin{align}\label{eq: profit}
   \text{Profit}= B_{\rm c} - \text{Cost}_{RT} -\text{Cost}_{DA}
\end{align} where $B_{\rm c}$ represents the total money paid by the customers for their electricity usage (by \eqref{eq:bill}, each customer will pay $\sum_{\ell=1}^Lp(\ell)  D_{{{\rm total}}}(\ell)$), and $\text{Cost}_{DA}$ denotes the cost for day-ahead market.
As discussed in the previous section, the HEMS ran in each residential unit will not only reduce the bill $B_{\rm c}$ but also potentially increase the real-time market cost $\text{Cost}_{RT}$ of the retailer. Hence it is desirable for the retailer to coordinate the HEMS of the residential units to reduce $\text{Cost}_{RT}$.

As incentives for the customers to participate in the coordinated HEMS program, we propose that 1) the retailer will charge the same amount of money from the customers as that optimized by their individual HEMS (i.e., \eqref{eq:HEMS}); 2) the coordinated HEMS will maintain the same scheduling deadline constraints specified by the customers. In summary, the customers would neither have any financial loss nor would lose any degree of comfort, if they joined the coordinated HEMS program. Nevertheless, the retailer will directly benefit from the reduction of $\text{Cost}_{RT}$ according to \eqref{eq: profit}.

\subsection{Proposed Coordinated HEMS}

Following the two conditions above, we propose to coordinate the scheduling tasks of the $M$ residential units, targeting at minimizing the real-time market cost in \eqref{eq:deviation cost}. To extend the load models in Section II-A to the $M$ residential units, we use superscript $(m)$ to denote the $m$th residential unit; for example, $t_{i,k}^{(m)}$ and $s_{i,k}^{(m)}$ represent the request arrival and task operating times of appliance $i$ in the $m$th residence, and $S^{(m)}_{i}(\ell)$ represents the controllable load injection of appliance $i$ in the $m$th residence.

The proposed coordinated HEMS design is given by
\begin{subequations}\label{eq:COHEMS}
\begin{align}
  \min_{s_{i,1}^{(m)},s_{i,2}^{(m)},\ldots}~&\sum_{\ell=1}^L \E\left[ \pi_{\rm s}(\ell)\left(E(\ell)-\sum_{m=1}^M D_{\rm total}^{(m)}(\ell)\right)^+ \notag \right.\\
    &\left.~~~~~~~+\pi_{\rm p}(\ell)\left(\sum_{m=1}^M D_{\rm total}^{(m)}(\ell)-E(\ell)\right)^+ \right] \\
  \text{s.t.}~& D_{\rm total}^{(m)}(\ell)=U^{(m)}(\ell) + \sum_{i=1}^N{S}_i^{(m)}(\ell),\label{eq:COHEMS b}\\
  & {S}_i^{(m)}(\ell)=\sum_{k=1}^{\infty} g_i^{(m)}(\ell- s_{i,k}^{(m)})~\forall~i,\ell,m, \label{eq:COHEMS c}\\
  & t_{i,k}^{(m)} \leq s_{i,k}^{(m)} \leq  t_{i,k}^{(m)}+\zeta_i^{(m)}~\forall~i,k,m, \label{eq:COHEMS d}\\
  & s_{i,k}^{(m)} \leq L~ \forall~i,k,m, \label{eq:COHEMS e}\\
  & \sum_{i=1}^N S_{i}^{(m)}(\ell) \leq P_{\max}~\forall~ \ell,m. \label{eq:COHEMS f}
\end{align}
\end{subequations}
Problem \eqref{eq:COHEMS} minimizes the average real-time market cost of the retailer. Note that problem \eqref{eq:COHEMS} is subject to the same scheduling constraints as problem \eqref{eq:HEMS} for each residential unit, meaning that the degree of comfort of customers is preserved in the proposed coordinated HEMS.

We show here that the coordinated HEMS problem \eqref{eq:COHEMS} can be expressed as a DP. The key observation is that finding the optimal operating times $s_{i,1}^{(m)}$, $s_{i,2}^{(m)}$, $\ldots$ is equivalent to finding the optimal task departure (launching) process $d_i^{(m)}(\ell)$. Specifically, in accordance with \eqref{eq:departure process}, the optimal $s_{i,k}^{(m)}$ is given by $\ell^\star$ if $\ell^\star$ is the minimum number in the set
$$
   \mathfrak{L}_{i}^{(m)}(k) = \{ \ell \in \{1,\ldots,L\} ~|~ d_i^{(m)}(\ell)=k\}.
$$
We should emphasize here that the observation above can significantly simplify the optimization of \eqref{eq:COHEMS}.

By the fact that the load injection in \eqref{eq:load injection of appliance i} is the convolution of the departure process difference $d_i^{(m)}(\ell)-d_i^{(m)}(\ell-1)$ and the load profile $g_i^{(m)}(\ell)$, one can express $S_i^{(m)}(\ell)$ in \eqref{eq:COHEMS c} as
\begin{align}\label{eq:departure process1}
  S_i^{(m)}(\ell) &=\sum_{k=1}^{\infty} [d_i^{(m)}(\ell-k+1)-d_i^{(m)}(\ell-k)]g_i^{(m)}(k) \notag \\
  & =\!\!\!\!\!\!\sum_{k=1}^{\min\{\ell,G_i^{(m)}\}} [d_i^{(m)}(\ell-k+1)-d_i^{(m)}(\ell-k)]g_i^{(m)}(k),
\end{align}where $d_i^{(m)}(0)=0$, and the second equality is owing to that $g_i^{(m)}(\ell)$ has a maximum duration $G_i^{(m)}$. Moreover, since $d_i^{(m)}(\ell)$ is nondecreasing and according to the scheduling constraints \eqref{eq:COHEMS d} and \eqref{eq:COHEMS e}, $d_i^{(m)}(\ell)$ should satisfy
\begin{subequations}\label{eq:departure process constraints}
\begin{align}
    d_i^{(m)}(\ell-1) \leq~  &d_i^{(m)}(\ell) \leq a_i^{(m)}(\ell), \\
    a_i^{(m)}(\ell-\zeta_i^{(m)})  \leq~  & d_i^{(m)}(\ell), \label{eq:departure process constraints b}\\
     &d_i^{(m)}(L)=a_i^{(m)}(L),\label{eq:departure process constraints c}\\
     &d_i^{(m)}(\ell) \in \mathbb{Z}_+,
\end{align}
\end{subequations} for all $\ell$, $i$ and $m$, where $\mathbb{Z}_+$ denotes the set of nonnegative integers. Specifically, \eqref{eq:departure process constraints b} guarantees that appliance $i$ will be scheduled within the maximum delay $\zeta_i^{(m)}$.

By \eqref{eq:departure process1} and \eqref{eq:departure process constraints}, we can reformulate problem \eqref{eq:COHEMS} as the following problem
\begin{subequations}\label{eq:COHEMS departure}
\begin{align}
\!\!  \min_{\substack{d_{i}^{(m)}(\ell)\\ \forall \ell,i,m}}&\sum_{\ell=1}^L \E\left[ \pi_{\rm s}(\ell)\left(\tilde{E}(\ell)-\sum_{m=1}^M \sum_{i=1}^N{S}_i^{(m)}(\ell)\right)^+ \notag \right.\\
    &\left.~~~~~~~+\pi_{\rm p}(\ell)\left(\sum_{m=1}^M \sum_{i=1}^N{S}_i^{(m)}(\ell)-\tilde{E}(\ell)\right)^+ \right] \label{eq:COHEMS departure a}\\
  \text{s.t.}~&  \sum_{i=1}^N S_{i}^{(m)}(\ell) \leq P_{\max}~\forall~ \ell,m, \label{eq:COHEMS departure b}\\
  & \text{constraints~in~} \eqref{eq:departure process1}-\eqref{eq:departure process constraints},
\end{align}
\end{subequations}
where $\tilde{E}(\ell)=E(\ell)-\sum_{m=1}^M U^{(m)}(\ell)$ (see \eqref{eq:total load}). Comparing with \eqref{eq:COHEMS}, in \eqref{eq:COHEMS departure},
the optimal departure processes $\{d_{i}^{(m)}(\ell)\}$ are to be determined instead.

Problem \eqref{eq:COHEMS departure} can be solved by the standard DP approach, e.g., using the principle of optimality of DP \cite{BK:Bersekas07}, by which the optimal control policy for $\{d_{i}^{(m)}(\ell)\}$ can be obtained in a backward search manner. This method, however, is not computationally feasible because \eqref{eq:COHEMS departure} involves a large dimension of state vector. In particular, the state vector corresponding to \eqref{eq:COHEMS departure} at stage $\ell$ is given by
\begin{align}
  \xb_\ell&=[\xb_{1,1}^T(\ell),\ldots,\xb_{1,N}^T(\ell),\xb_{2,1}^T,\ldots,\xb_{M,N}^T(\ell)]^T, \end{align}where
\begin{align}
 \xb_{m,i}(\ell)&=[d_i^{(m)}(\ell-1),\ldots,d_i^{(m)}(\ell-\min\{\ell,G_i^{(m)}\}),\notag\\
   &~~~~~~~~~~~~~~~~~a_i^{(m)}(\ell),a_i^{(m)}(\ell-\zeta_i^{(m)})]^T.
\end{align} As seen, the number of possibilities of $\xb_\ell$ exponentially increase with $M$ and $N$.

\subsection{Certainty Equivalent Control (CEC)}
Certainty Equivalent Control (CEC) is a simple approach to obtaining an approximate solution of a complicated DP problem \cite{BK:Bersekas07}. In CEC, we search the optimal control in a forward manner and apply the control at each time that would be optimal if the uncertainty quantities were fixed at the typical values, e.g., the mean value. Therefore, by CEC, we can obtain an approximate solution to \eqref{eq:COHEMS departure} in an \emph{on-line} fashion, sequentially from time $1$ to $L$, and each time we only need to deal with a deterministic optimization problem. Applying CEC to problem \eqref{eq:COHEMS departure}, we obtain the following algorithm:

\begin{algorithm}[h]
  \caption{CEC approach to problem \eqref{eq:COHEMS departure}}
\begin{algorithmic}[1]

    \STATE \text{\bf for}~\text{time}~{$\bar \ell=1,\dots,L-1$}~\text{\bf do}
      \STATE Given $\xb_{\bar\ell}$, solve the following problem
      {\small
        \begin{align}\label{eq:COHEMS departure cec}
        \!\!\!\!\!\!  \min_{\substack{d_{i}^{(m)}(\ell)~\forall i,m\\\ell=\bar\ell,\ldots,L }}&\sum_{\ell=\bar \ell}^L \left[ \pi_{\rm s}(\ell)\left(\tilde{E}(\ell)-\sum_{m=1}^M \sum_{i=1}^N{S}_i^{(m)}(\ell)\right)^+ \notag \right.\\
            &\left.~~~~~+\pi_{\rm p}(\ell)\left(\sum_{m=1}^M \sum_{i=1}^N{S}_i^{(m)}(\ell)-\tilde{E}(\ell)\right)^+\!\! \right] \\
          \text{s.t.}~&  \sum_{i=1}^N S_{i}^{(m)}(\ell) \leq P_{\max}, \notag \\
          & d_i^{(m)}(\ell-1) \leq  d_i^{(m)}(\ell) \leq a_i^{(m)}(\bar \ell)+\sum_{k=\bar \ell+1}^\ell \alpha_{i}^{(m)}(k) \notag \\
          &  \Phi_i^{(m)}(\ell)  \leq   d_i^{(m)}(\ell), \notag\\
             &d_i^{(m)}(L)=a_i^{(m)}(\bar \ell)+\sum_{k=\bar \ell+1}^L \alpha_{i}^{(m)}(k), \notag \\
             &d_i^{(m)}(\ell) \in \mathbb{Z}_+, ~\forall~ m,i,\ell=\bar \ell,\ldots,L, \notag
        \end{align}
        }
        \!\!and denote $\{\bar{d}_{i}^{(m)}(\ell)\}_{i,m,\ell}$ as the associated optimal solution.
    \STATE \text{\bf Set} ${d}_{i}^{(m)}(\bar \ell)=\bar{d}_{i}^{(m)}(\bar \ell)$ for all $i,m$, as the approximate solution at time $\bar \ell$.

    \STATE \text{\bf end~for}
\end{algorithmic}
\end{algorithm}
\noindent In \eqref{eq:COHEMS departure cec}, ${S}_i^{(m)}(\ell)$ is given by \eqref{eq:departure process1}, and
$\Phi_i^{(m)}(\ell)$ is defined as
\begin{align*}
  \Phi_i^{(m)}(\ell)=\!\!\left\{\!\!\!\begin{array}{ll}
                               a_i^{(m)}(\ell-\zeta_i^{(m)}),&\ell=\bar \ell,\ldots,\bar\ell+\zeta_i^{{(m)}} \\
                               a_i^{(m)}(\bar\ell)+\!\!\sum_{k=\bar\ell+1}^{\ell-\zeta_i^{(m)}}
                               \alpha_i^{(m)}(k),&
                               \text{elsewhere.}
                               \end{array}\right.
\end{align*}
Note that, in \eqref{eq:COHEMS departure cec}, the unknown arrivals $a_i^{(m)}(\ell),~\ell=\bar\ell+1,\ldots,L$, at time $\bar\ell$ are set to their mean values $a_i^{(m)}(\bar \ell)+\sum_{k=\bar \ell+1}^\ell \alpha_{i}^{(m)}(k),~\ell=\bar\ell+1,\ldots,L$, so problem \eqref{eq:COHEMS departure cec} is a deterministic optimization problem for all $\bar\ell=1,\ldots,L$.

Problem \eqref{eq:COHEMS departure cec} has a convex objective function and convex constraints, except for the integer constraints $d_i^{(m)}(\ell) \in \mathbb{Z}_+$. Since the integer constraints lead to a discrete optimization problem which is difficult to handle in general, we simply relax the integer constraints to nonnegative orthant $d_i^{(m)}(\ell)\geq 0$. An approximate solution to \eqref{eq:COHEMS departure cec} can be obtained by rounding the solutions of the relaxed problem into the nearest integers.
Next we show that the relaxed counterpart of problem \eqref{eq:COHEMS departure cec} can be recast as a linear programming (LP) which thus can be solved efficiently. To illustrate this, let us first express \eqref{eq:COHEMS departure cec} in a compact form. Define
\begin{align*}
\!\!\!\!\pib_{\rm p}=&[\pi_{\rm p}(L),\ldots,\pi_{\rm p}(\bar\ell)]^T, \notag \\
\pib_{\rm s}=&[\pi_{\rm s}(L),\ldots,\pi_{\rm s}(\bar\ell)]^T, \\
  \tilde{\Eb}=&[\tilde{E}(L),\ldots,\tilde{E}(\bar\ell)]^T, \\
  \tilde\db^{(m)}=&[ d_1^{(m)}(L),\ldots,d_1^{(m)}(\bar\ell-\min\{\bar\ell,G_1^{(m)}\}), \notag \\
  &~~~~~~~~~~d_2^{(m)}(L),\ldots,d_N^{(m)}(\bar\ell-\min\{\bar\ell,G_N^{(m)}\})]^T, \\
  \db^{(m)}=&[ d_1^{(m)}(L),\ldots,d_1^{(m)}(\bar\ell), d_2^{(m)}(L),\ldots,d_N^{(m)}(\bar\ell)]^T, \\
  \Psib^{(m)} =&[\Omegab_1^{(m)},\ldots,\Omegab_N^{(m)}]\text{blkdiag}\{\Upsilonb_1^{(m)},\ldots,\Upsilonb_N^{(m)}\},
\end{align*}where $\Omegab_i^{(m)}\in \mathbb{R}^{(L-\bar\ell+1)\times(L-\bar\ell+\min\{\bar\ell,G_i^{(m)}\})}$ is a Toeplitz matrix with the first row given by $[g_i^{(m)}(1),\ldots,g_i^{(m)}(G_i^{(m)})]$ and the first column given by $[g_i^{(m)}(1),0,\ldots,0]^T$, and $\text{blkdiag}\{\Upsilonb_1^{(m)},\ldots,\Upsilonb_N^{(m)}\}$ is a block diagonal matrix in which $\Upsilonb_i^{(m)} \in \mathbb{R}^{(L-\bar\ell+\min\{\bar\ell,G_i^{(m)}\})\times(L-\bar\ell+\min\{\bar\ell,G_i^{(m)}\}+1)}$
is a Toeplitz matrix with the first row being $[1, -1, 0, \ldots, 0]$ and the first column being $[1, 0, \ldots, 0]^T$. Moreover, define
\begin{align}
   &\mathcal{U}^{(m)}=\bigg\{ \db^{(m)}\succeq \zerob |~  \Psib^{(m)} \tilde\db^{(m)} \preceq  P_{\max} {\bf 1}, \bigg.\notag\\
   &\!\!\!\!\left.\begin{array}{ll}
   &d_i^{(m)}(\ell-1) \leq  d_i^{(m)}(\ell) \leq a_i^{(m)}(\bar \ell)+\sum_{k=\bar \ell+1}^\ell \alpha_{i}^{(m)}(k), \\
   &\Phi_i^{(m)}(\ell)  \leq   d_i^{(m)}(\ell),\\
   &d_i^{(m)}(L)=a_i^{(m)}(\bar \ell)+\sum_{k=\bar \ell+1}^L \alpha_{i}^{(m)}(k)~
   \forall~ i,\ell=\bar \ell,\ldots,L, \\
   \end{array}\!\!\!\!\!
   \notag \right\},
\end{align}
where $\preceq$ and $\succeq$ denote the element-wise inequalities, and ${\bf 1}$ ($\zerob$) is the all-one (all-zero) vector.
Then problem \eqref{eq:COHEMS departure cec} can be expressed as
\begin{align}\label{eq:COHEMS departure cec2}
        \!\!\!\!\!\!  \min_{\substack{\db^{(m)}\\m=1,\ldots,M }}& \pib_{\rm s}^T\left(\tilde{\Eb}-\sum_{m=1}^M \Psib^{(m)} \tilde\db^{(m)}\right)^+ \notag \\
            &~~~~~~~~~~~~+\pib_{\rm p}^T\left(\sum_{m=1}^M \Psib^{(m)} \tilde\db^{(m)}-\tilde{\Eb}\right)^+\!\!\\
          \text{s.t.}~&  \db^{(m)} \in \mathcal{U}^{(m)},~m=1,\ldots,M.
\end{align}

By introducing a slack variable
\begin{align}\label{eq:z}\zb=\left(\sum_{m=1}^M \Psib^{(m)} \tilde\db^{(m)}-\tilde{\Eb}\right)^+,\end{align} one can write
\begin{align*}
\left(\tilde{\Eb}-\sum_{m=1}^M \Psib^{(m)} \tilde\db^{(m)}\right)^+=\zb-
\left(\sum_{m=1}^M \Psib^{(m)} \tilde\db^{(m)}-\tilde{\Eb}\right).
\end{align*}
Substituting the above equations into \eqref{eq:COHEMS departure cec2} gives rise to
\begin{subequations}\label{eq:COHEMS departure cec LP0}
\begin{align}
        \!\!\!\!\!\!  \min_{\substack{d_{i}^{(m)}(\ell)~\forall i,m\\\ell=\bar\ell,\ldots,L \\
        \zb \in \mathbb{R}^{L-\bar\ell+1} }}& (\pib_{\rm s}+\pib_{\rm p})^T\zb-\pib_{\rm s}^T\left(\sum_{m=1}^M \Psib^{(m)} \tilde\db^{(m)}-\tilde{\Eb}\right)\notag\\
          \text{s.t.}~&  \zb=\left(\sum_{m=1}^M \Psib^{(m)} \tilde\db^{(m)}-\tilde{\Eb}\right)^+, \label{eq:COHEMS departure cec LP0 a} \\
          &\db^{(m)} \in \mathcal{U}^{(m)},~m=1,\ldots,M. \notag
\end{align}
\end{subequations}
Assume the usual case of $\pib_{\rm s}+\pib_{\rm p}\succeq \zerob$. Then the constraint \eqref{eq:COHEMS departure cec LP0 a} can be shown to be equivalent to the following two linear constrains:
\begin{align}\label{eq:z2}
  \zb \succeq \zerob,~\zb \succeq \sum_{m=1}^M \Psib^{(m)} \tilde\db^{(m)}-\tilde{\Eb}.
\end{align}
By replacing \eqref{eq:COHEMS departure cec LP0 a} with \eqref{eq:z2}, we end up with the following LP representation for \eqref{eq:COHEMS departure cec}:
\begin{subequations}\label{eq:COHEMS departure cec LP}
\begin{align}
        \!\!\!\!\!\!  \min_{\substack{d_{i}^{(m)}(\ell)~\forall i,m\\\ell=\bar\ell,\ldots,L \\
        \zb \in \mathbb{R}^{L-\bar\ell+1} }}& (\pib_{\rm s}+\pib_{\rm p})^T\zb-\pib_{\rm s}^T\left(\sum_{m=1}^M \Psib^{(m)} \tilde\db^{(m)}-\tilde{\Eb}\right)\notag\\
          \text{s.t.}~&  \zb \succeq \zerob, \label{eq:COHEMS departure cec LP a} \\
          &\zb \succeq \sum_{m=1}^M \Psib^{(m)} \tilde\db^{(m)}-\tilde{\Eb}, \label{eq:COHEMS departure cec LP b} \\
          &\db^{(m)} \in \mathcal{U}^{(m)},~m=1,\ldots,M. \notag
\end{align}
\end{subequations}

As a remark, we should mention that the reformulation idea in Section III-B and the CEC method in Section III-C can also be applied for handling the individual HEMS problem in \eqref{eq:HEMS}.

\section{Distributed Implementation}
In the previous section, we have shown how the proposed coordinated HEMS problem \eqref{eq:COHEMS} can be approximated by CEC which involves only solving the integer-constraint-relaxed LP problem \eqref{eq:COHEMS departure cec LP}. To solve problem \eqref{eq:COHEMS departure cec LP}, a centralized control is needed in general. The control center not only knows the appliance profiles in each residential units, but also the statistical and real-time information of the request arrival processes $\{a_i^{(m)}(\ell)\}$. In view of the fact that the computational complexity of solving
\eqref{eq:COHEMS departure cec LP} increases with the number of residences and the number of controllable appliances, a decentralized implementation algorithm, that can decompose the original problem into parallel subproblems with smaller problem size, is of great interest. In particular, we are interested in decentralized algorithms that allow each of the residential units to compute its scheduling solution locally using only domestic information so that the customers' privacy on electricity usage can also be preserved.

In this section, we present a decentralized implementation method for problem \eqref{eq:COHEMS departure cec LP}, using the convex optimization based dual decomposition method \cite{Boyddecomposition}. Combining such a decentralized method with Algorithm 1, a simple distributed coordinated HEMS algorithm is obtained.

As its name suggests, dual decomposition solves the problem in the Lagrangian dual domain. Let $\mub \succeq \zerob$ and $\lambdab \succeq \zerob$ be the dual variables associated with the inequality constraints in \eqref{eq:COHEMS departure cec LP a} and \eqref{eq:COHEMS departure cec LP b}, respectively. By definition \cite{BK:Boyd04}, the Lagrangian dual problem of \eqref{eq:COHEMS departure cec LP} can be shown to be
\begin{align}\label{eq: dual problem}
  \max_{\mub \succeq \zerob,\lambdab \succeq \zerob}~\phi(\mub,\lambdab)
\end{align}
where $\phi(\mub,\lambdab)$ is the dual function given by
\begin{align*}
        \!\!\!\!\!\!  &\min_{\substack{\db^{(m)} \in \mathcal{U}^{(m)}~\forall m,\\
        \zb \in \mathbb{R}^{L-\bar\ell+1} }} (\pib_{\rm s}+\pib_{\rm p})^T\zb-\pib_{\rm s}^T\left(\sum_{m=1}^M \Psib^{(m)} \tilde\db^{(m)}-\tilde{\Eb}\right)\notag\\
        &~~~~~~~~~~~~~~~~~~-\mub^T\zb+\lambdab^T\left(\sum_{m=1}^M \Psib^{(m)} \tilde\db^{(m)}-\tilde{\Eb}-\zb\right) \notag \\
        &=\left\{\!\!\!\!\!
        \begin{array}{ll}
        &{\displaystyle \min_{\substack{\db^{(m)} \in \mathcal{U}^{(m)}~\forall m }} (\lambdab-\pib_{\rm s})^T\left(\sum_{m=1}^M \Psib^{(m)} \tilde\db^{(m)}-\tilde{\Eb}\right)
        }\\
        &~~~~~~~~~~~~~~~~~~~~~~~~~~~~~~~~~~~\text{if}~\pib_{\rm p}+\pib_{\rm s}-\lambdab=\mub, \\
        &-\infty,~\text{elsewhere}.
        \end{array}
        \right.
\end{align*}
Substituting the above equation into \eqref{eq: dual problem} gives rise to
\begin{align}\label{eq: dual problem2}
  \max_{ \zerob \preceq \lambdab \preceq  \pib_{\rm p}+\pib_{\rm s}}~
  \!\!\!\!\left\{
  \!\!\!\!\!\!\!\!\begin{array}{ll}
  &{\displaystyle\min_{\substack{\db^{(m)}~\forall m }}~ (\lambdab-\pib_{\rm s})^T\left(\sum_{m=1}^M \Psib^{(m)} \tilde\db^{(m)}-\tilde{\Eb}\right)} \\
  &~~~~~~~~~ {\displaystyle \text{s.t.}~\db^{(m)} \in \mathcal{U}^{(m)},~ m=1,\ldots,M.}
        \end{array}\!\!\!\!\!
  \right\}
\end{align}
Since problem \eqref{eq:COHEMS departure cec LP} is convex and satisfies Slater's condition \cite{BK:Boyd04}, the dual problem \eqref{eq: dual problem2} attains the same optimal objective value as \eqref{eq:COHEMS departure cec LP}.
One can see from \eqref{eq: dual problem2} that the inner part is decomposible, and can be solved in a parallel fashion given $\lambdab$.
Therefore, a decentralized implementation can be obtained by solving the dual problem \eqref{eq: dual problem2}. Specifically, we can use the projected subgradient method \cite{Boydsubgradient} to deal with \eqref{eq: dual problem2} in an iterative manner. In iteration $n$, given $\lambdab(n)$,
the corresponding inner part of \eqref{eq: dual problem2} can be handled by solving:
\begin{align}\label{eq:inner minimization2}
 \db^{(m)}(n+1)=\arg~\min_{\substack{\db^{(m)}\in \mathcal{U}^{(m)}}}~ (\lambdab(n)-\pib_{\rm s})^T\Psib^{(m)} \tilde\db^{(m)},
\end{align} for $m=1,\ldots,M$. The dual variable $\lambdab$ can be updated using the standard subgradient step \cite{Boydsubgradient}, i.e.,
\begin{align}\label{eq:dual update}
  \lambdab(n+1)\!\!= \mathcal{P}\left(\lambdab(n) + c_n \left(\sum_{m=1}^M \Psib^{(m)} \tilde\db^{(m)}(n+1)-\tilde{\Eb}\right)\right),
\end{align}
where $c_n>0$ denotes the step size, and $\mathcal{P}(\cdot)$ denotes the operation of projection onto the set $[\zerob, \pib_{\rm p}+\pib_{\rm s}]$. Equations \eqref{eq:dual update} and \eqref{eq:inner minimization2} are iterated until convergence or the preset stopping criterion is satisfied. The dual decomposition method for \eqref{eq:COHEMS departure cec LP} is summarized in Algorithm 2. Suppose that the algorithm stops at iteration $n^\star$. Instead of using $\{\db^{(m)}(n^\star)\}$ as the primal solution, we use the
\emph{running-averaged} version:
\begin{align}\label{eq:running average}
  &\hat\db^{(m)}=\frac{1}{n^\star+1}\sum_{q=1}^{n^\star+1}\db^{(m)}(q),~m=1,\ldots,M.
\end{align}
It is shown \cite{Angelia2009} that this averaged version $\hat\db^{(m)}$ is more numerically stable than $\db^{(m)}(n^\star)$, especially for our problem \eqref{eq:COHEMS departure cec LP} which is not strictly convex.

In Algorithm 2, we assume that there is a control center which collects $\db^{(m)}(n+1)$ from the residences and uses the information for updating the dual variable $\lambdab$ (Step 7). 
If there is no control center present, the residences can still perform subgradient update \eqref{eq:dual update} individually by obtaining the aggregate load profile $\sum_{m=1}^M \Psib^{(m)} \tilde\ssb^{(m)}(n+1))$ in a fully distributed fashion, e.g., using the average consensus algorithms or gossip algorithms
\cite{Xiao2003,Boyd2006}.
%
%

\begin{algorithm}[t]
  \caption{{Dual decomposition for problem \eqref{eq:COHEMS departure cec LP}}}
\begin{algorithmic}[1]\label{alg:decentralized_modified}
  \STATE {\bf Input} an initial value of $\lambdab(0)$.
  \STATE Set $n=0$
  \REPEAT
    \FOR{$m=1,\dots, M$}
      \STATE Residence $m$ solves \eqref{eq:inner minimization2} to
             obtain the solution $\db^{(m)}(n+1)$, and sends it to the control center.
    \ENDFOR
    \STATE Given $\db^{(m)}(n+1)$, $m=1,\ldots,M$, the control center updates the dual variable $\lambdab$ by \eqref{eq:dual update}, and broadcasts it to the residences.
   \STATE $n=n+1$
  \UNTIL the predefined stopping criterion is met.
\end{algorithmic}
\end{algorithm}

\section{Simulation Results}

In this section, some simulation results are presented to examine the effectiveness of the proposed coordinated HEMS. We consider a scenario where there are $60$ residential units ($M=60$), with 3 controllable appliances in each residence ($N=3$). The optimization horizon is set to $96$ ($L=96$) which is obtained by considering a whole day with 24 hours and 4 quarters for each hour (starting from 8 pm to the next day). The three appliances are assumed to have rectangular power profiles, with instantaneous energy consumptions uniformly generated between $[0.8,1.9]$ (kWh) (e.g., PHEV), $[0.3,0.5]$ (e.g., dish washer) and $[0.8,1.2]$ (e.g., cloth dryer), respectively (reference from {http://www.absak.com/library/power-consumption-table}). The simulation setting of
$\{G_i^{(m)}\}$ and $\{\zeta_{i}^{(m)}\}$ are detailed in Table 1. We assume that each residence will send a request for Appliance 1 with probability 0.8 in a time uniformly distributed between 8 pm and midnight, and with probability 0.3 in a time uniformly distributed between 8 am and 12 pm. Appliance 2 is set to probability $0.8$ in three times that are uniformly distributed between 6 am and 10 am, 12 pm and 2 pm, and 5 pm and 7 pm, respectively. Appliance 3 is set to probabilities $0.8$ and $1$ in the times between 2 pm and 3 pm, and 8 pm and 10 pm, respectively.

\begin{table}[t]\centering \vspace{-0.3cm}
\caption{Simulation Setting of Appliances. The notion $U\sim[a,b]$ stands for a uniform distribution in the interval $[a,b]$.
}\label{table:ex1margin}\vspace{-0.2cm}
\begin{center}
\begin{tabular}{>{\rm}c|>{\rm}c|>{\rm}c|>{\rm}c}
                & Appliance 1   & Appliance 2   &  Appliance 3 \\ \hline
  $g_i^{{m}}(\ell)$  (kWh) & $U\sim[3.25,7.5]$    & $U\sim[1.2,1.5]$  & $U\sim[0.3,0.5]$\\ \hline
  $G_i^{(m)}$  (quarter)       & $U\sim[16,32]$   & $U\sim[2,4]$  & $U\sim[4,12]$ \\ \hline
  $\zeta_{i}^{(m)}$ (quarter)  & $U\sim[4,16]$   &   $U\sim[4,12]$ & $U\sim[4,12]$   \\ \hline
\end{tabular}
\end{center}
\vspace{-0.0cm}
\end{table}

\begin{figure}[!t] \centering
   \resizebox{0.470\textwidth}{!}{
     \psfrag{gamma}[Bl][Bl]{\huge $\gamma$ (dB)}
      \includegraphics{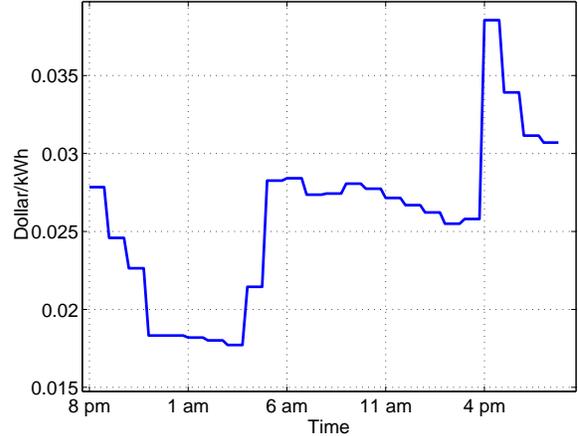}}
      \vspace{-0.0cm}
      \caption{Day ahead price obtained from https://www2.ameren.com/RetailEnergy/realtimeprices.aspx, on day Nov. 15, 2001.}
     \label{fig. price}\vspace{-0.3cm}
\end{figure}

For HEMS in \eqref{eq:HEMS}, we use the day-ahead price as shown in Figure \ref{fig. price}. For the proposed coordinated HEMS in \eqref{eq:COHEMS}, we consider two examples. In the first example, we set $\pib_{\rm p}=\pib_{\rm s}={\bf 1}$, by which the objective value of \eqref{eq:COHEMS} reduces to $\sum_{\ell=1}^L |E(\ell)-\sum_{m=1}^M D_{\rm total}^{(m)}(\ell)|$. We use this setting to examine the deviation between the scheduled load and the energy supply. In the second example, we set $\pib_{\rm p}={\bf 1}$ and $\pib_{\rm s}=-0.5{\bf 1}$, simulating the scenario that the retailer is able to sell the extra electricity back to the grid. In the simulations, for simplicity, we assume that, in each residence, the uncontrollable appliances contribute a constant instantaneous energy consumption of 5 kWh, which is assumed to be known by the residence in advance by prediction.

\begin{figure}[!t] \centering
   \resizebox{0.50\textwidth}{!}{
     \psfrag{gamma}[Bl][Bl]{\huge $\gamma$ (dB)}
      \includegraphics{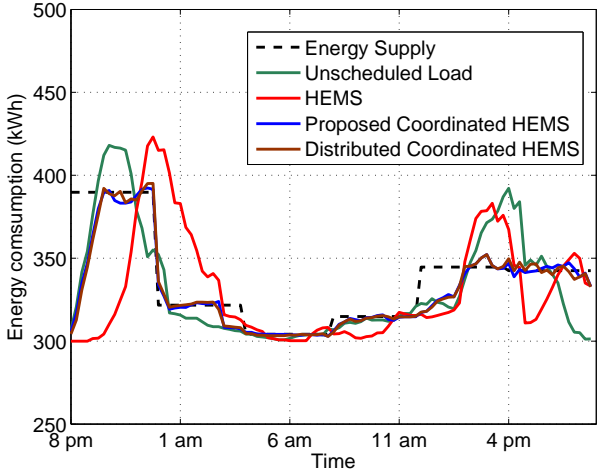}}
      \vspace{-0.3cm}
      \caption{Simulation results for a randomly generated problem instance with $\pib_{\rm p}=\pib_{\rm s}={\bf 1}$.}
     \label{fig. case1}\vspace{-0.0cm}
\end{figure}

\begin{figure}[!t] \centering
   \resizebox{0.50\textwidth}{!}{
     \psfrag{gamma}[Bl][Bl]{\huge $\gamma$ (dB)}
      \includegraphics{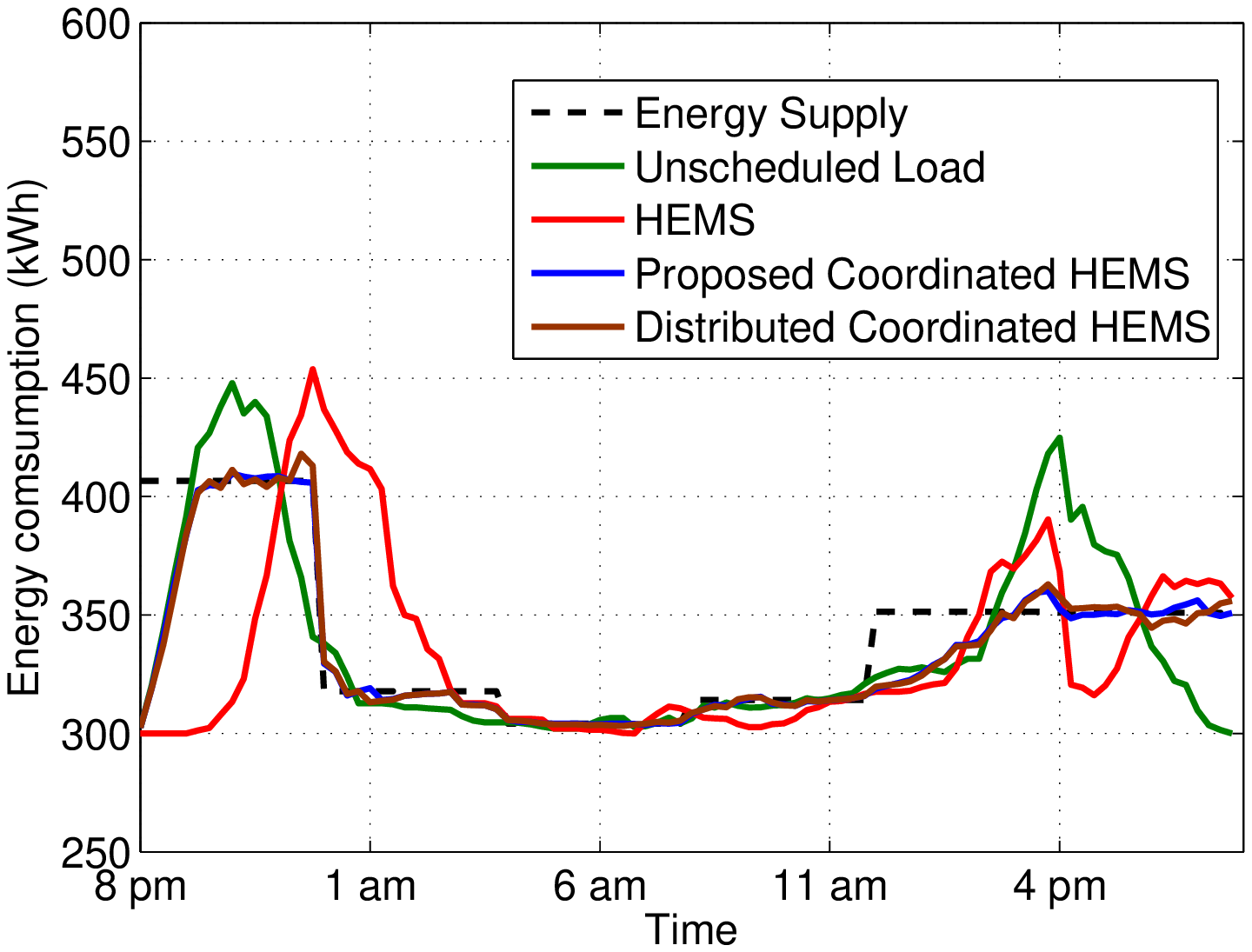}}
      \vspace{-0.3cm}
      \caption{Simulation results for a randomly generated problem instance with $\pib_{\rm p}={\bf 1}$ and $\pib_{\rm s}=-0.5{\bf 1}$.}
     \label{fig. case2}\vspace{-0.0cm}
\end{figure}

{\bf Example 1:} Figure \ref{fig. case1} shows the simulation results for a randomly generated problem instance with $\pib_{\rm p}=\pib_{\rm s}={\bf 1}$. Firstly, one can see from this figure and Figure \ref{fig. price} that the HEMS (i.e., \eqref{eq:HEMS}) successfully move the load to the lower price region, but that causes significant power imbalance. Specifically, the deviation $\sum_{\ell=1}^L |E(\ell)-\sum_{m=1}^M D_{\rm total}^{(m)}|$ corresponding to the unscheduled load is 1494, but the deviation corresponding to individual HEMS increases to 2450. Secondly, we can see from Figure \ref{fig. case1} that the proposed coordinated HEMS can schedule the load such that the corresponding load follows the energy supply. The load deviation of the proposed coordinated HEMS dramatically decreases to 698. Thirdly, we can observe that the distributed coordinated HEMS can yield almost the same performance as its centralized counterpart.

{\bf Example 2:} Figure \ref{fig. case2} shows the simulation results for another randomly generated problem instance with $\pib_{\rm p}={\bf 1}$ and $\pib_{\rm s}=-0.5{\bf 1}$. The simulation results are similar to Figure \ref{fig. case1}. In this case, the real-time cost in \eqref{eq:deviation cost} corresponding to the unscheduled load is 139 and that corresponding to HEMS is 276. The proposed coordinated HEMS however can reduce the cost to -246. This demonstrates well the efficacy of the proposed coordinated HEMS.

\vspace{-0.0cm}
\section{Conclusions}
In the paper, we have presented a coordinated HEMS architecture that coordinates the home energy scheduling of multiple residential units in order to reduce the real-time market cost of the retailer. We have shown that the coordinated HEMS design problem can be reformulated as a DP which can be efficiently handled by CEC. Moreover, a distributed implementation method by dual decomposition is also proposed. The presented Simulation results have shown that the proposed coordinated HEMS can effectively achieve real-time power balancing in contrast to the individual HEMS that may cause a rebound peak load in the low-price region.

\vspace{-0.0cm}
\section{Acknowledgements}
This work is supported in part by the US Department of Energy under the Trustworthy Cyber Infrastructure for the Power Grid (TCIPG) program.
\vspace{-0.0cm}
\vspace{-0.0cm} \footnotesize
\bibliography{smart_grid}
\end{document}